\documentclass[aps,prd,groupedaddress,showpacs,graphicx,nofootinbib]{revtex4}
\usepackage{amssymb,graphics,graphicx,color}
\def\lae{\;^{<}_{\sim} \;} \def\gae{\; ^{>}_{\sim} \;}
\begin{document}

\title{Cosmon as the Modulon: Non-Gaussianity from Dark Energy}
\author{Chian-Shu Chen$^{1,3}$\footnote{chianshu@phys.sinica.edu.tw} and Chia-Min Lin$^{2,3}$\footnote{cmlin@phys.nthu.edu.tw}}

\affiliation{$^{1}$Physics Division, National Center for Theoretical Sciences, Hsinchu, Taiwan 300\\$^{2}$Department of Physics, National Tsing Hua University, Hsinchu, Taiwan 300\\$^{3}$Institute of Physics, Academia Sinica, Taipei, Taiwan 115}
\date{Draft \today}

\begin{abstract}
In this paper, we show that the idea of growing neutrino can result in a modulated reheating effect and produce detectable non-Gaussianity in the model where the Higgs triplet from type II seesaw mechanism plays the role of the inflaton in chaotic inflation.
\end{abstract}
\maketitle

\underline{\emph{Introduction}}:
Inflation \cite{Lyth:2009zz} not only solves many problems (for example, horizon problem and flatness problem) of conventional hot big bang model but could also produce the primordial density perturbation which is the seed for structure formation and cosmic microwave background (CMB) temperature fluctuations. Intuitively, the mechanism for producing density fluctuation from inflation is that during inflation the quantum fluctuations of the inflaton when stretched outside the horizon becomes classical perturbations which is different in different patches of the universe separated by the horizon. Each patch could be regarded as a ``separate universe'' and evolves in the same way. However, inflation ends in different ``time'' for each universe and this result in a primordial density (curvature) perturbation. The primordial curvature perturbation $\zeta$ (on uniform-density slices) is given by (with a suitable coordinate choice)
\begin{equation}
dl^2=a^2(t)e^{2\zeta}dx^idx^j \sim a^2(t)(1+2\zeta)dx^idx^j.
\end{equation}
Here $\zeta$ describes the difference between the perturbed universe and unperturbed universe. During inflation, since $a \sim e^{N}$, therfore intuitively one may think $\delta N = \zeta$. This has been rigorously proved and is called the $\delta N$ formalism \cite{Sasaki:1995aw, Sasaki:1998ug, Lyth:2004gb, Lyth:2005fi}. This relation is true up to nonlinear orders. For example, if the primordial density perturbation is from the fluctuation of the inflaton $\phi$, we can write
\begin{equation}
\delta N=N_\phi \delta \phi + \frac{1}{2}N_{\phi\phi} (\delta\phi)^2+\cdots
\label{eq2}
\end{equation}
where subscript denotes derivative with respective to the corresponding argument. Since $\delta \phi \sim H/ 2\pi$ is Gaussian, the second term represents non-Gaussianity.
Future experiments (like PLANCK) can probe the second (or even higher) order effects for the primordial curvature perturbation which is parameterized as
\begin{equation}
\delta N =\zeta=\zeta_g+\frac{3}{5}f_{NL} \zeta_g^2+\cdots.
\label{eq3}
\end{equation}
Here $\zeta_g$ is the linear (Gaussian) term and $f_{NL}$ is the (local) nonlinear parameter.
From Eqs.~(\ref{eq2}) and (\ref{eq3}), we could obtain
\begin{equation}
f_{NL}=\frac{5}{6}\frac{N_{\phi\phi}}{N_{\phi}^2}.
\label{eq4}
\end{equation}
Generally, for single-field slow-roll inflation, $f_{NL}$ is too small to be detected in the near future \cite{Maldacena:2002vr}. Therefore a detection will force us to go for more complicated (inflation) models.

In Ref.~\cite{Chen:2010uc}, we proposed the idea that the Higgs triplet $\Delta$ in type II seesaw mechanism could play the role of the inflaton for chaotic inflation. The potential form is
\begin{equation}\label{inflatonpotential}
V=\frac{1}{2}M^2_{\Delta} \Delta^2,
\label{eq5}
\end{equation}
where the quartic terms are ignored\footnote{There is an argument given in \cite{Linde:2007fr} about why we may neglect the quartic terms.}. The number of e-folds is hence given by
\begin{eqnarray}
N&=&\frac{1}{M^2_P} \int \frac{V}{V^{\prime}} d\phi \nonumber \\
 &=&\frac{1}{M^2_P}\frac{\phi^2}{4}
\label{eq6}
\end{eqnarray}
which is independent of the mass. Here $M_P$ is the reduced Planck mass. From Eq.~(\ref{eq2}) and (\ref{eq6}), we can see $\zeta \sim H_\ast \sim {\cal O}(10)M_{\Delta}$ which is fixed to be ${\cal O}(10^{-5})$ by CMB observation. Here $H_\ast$ means the Hubble parameter at horizon exit.
If the primordial density perturbation is from the quantum fluctuation of the inflaton, this would imply that the inflaton mass $m_\Delta$ is fixed to be around $10^{13}$ GeV and the primordial density perturbation would be gaussian. However, it is possible that the primordial density perturbation is from some other mechanism and we could have $H_\ast \lae 10^{-5}$. For example, in the case of modulated reheating scenario  \cite{Dvali:2003em}, the inflaton decay width is determined by some light field $\sigma$ (called the modulon). During inflation the quantum fluctuation of the modulon will ``modulate'' the decay width of the inflaton hence when inflation decays after inflation this will contribute to the primordial density perturbation. The number of e-folds is related to the decay width $\Gamma$ via
\begin{equation}
N=-\frac{1}{6}\ln \Gamma.
\end{equation}
In this case if the contribution of $\delta N$ from inflaton is negaligible, similar to Eq.~(\ref{eq2}) we could write
\begin{equation}
\delta N=N_\sigma \delta \sigma + \frac{1}{2}N_{\sigma\sigma} (\delta\sigma)^2+\cdots.
\end{equation}
Therefore to first order we obtain
\begin{equation}
\delta N = -\frac{1}{6}\frac{\delta \Gamma}{\Gamma}
\label{eq7}
\end{equation}
and similar to Eq.~(\ref{eq4}), we can easily obtain
\begin{equation}
\frac{6}{5}f_{NL}=6 \left( 1-\frac{\Gamma \Gamma_{\sigma\sigma}}{\Gamma^2_{\sigma}} \right).
\label{eq10}
\end{equation}

One possibility of the dependence of the decay width is that the \emph{inflaton mass} is determined by the field value of the modulon. For a Higgs triplet, this kind of dependence was proposed in Ref.~\cite{Amendola:2007yx,Wetterich:2007kr} for another purpose and the light field is called the cosmon. In this model the cosmon field would make the neutrino mass growing (growing neutrino) through the varying mass of the Higgs triplet via type II seesaw mechanism. The result is in the current universe the cosmon field will be freezed due to growing neutrino and the scalar potential of the cosmon would become the dark energy we observe today. In this paper, we will show that cosmon field in the early universe could play the role of the modulon in the Higgs triplet inflation\footnote{We study the possible role of the cosmon in the early universe in a different set up in \cite{Chen:2011dv}.} and produce detectable primordial non-Gaussianity. The constraint to the Higgs triplet mass can also be liberated. The seesaw conception used in particle physics to understand the smallness of neutrino mass through the high energy physics can apply to connect the early universe (inflation $\sim$ Grand Unification scale) and current universe (dark energy $\sim$ eV scale) in our model. 

\underline{\emph{Construction}}: 
The type-II seesaw mechanism contains one complex $SU(2)_{L}$ triplet scalar $\Delta$ with hypercharge $Y = 2$ in addition to the standard model Higgs doublet $H$~\cite{type-II}. The Higgs triplet $\Delta$ can interact with left-handed leptons through the coupling, $Y_{ij}L^{T}_{iL}Ci\tau_{2}\Delta L_{jL}$ + H.c., here $i$ is the flavor index, $C$ is the charge conjugation, and $\tau_{2}$ is the Pauli matrix. There is also a trilinear term, $\mu H^{T}i\tau_{2}\Delta^{\dag}H$ + H.c. in the potential ($\mu$ is the dimensionful parameter). The coexistence of both two terms breaks the lepton number by two units. After taking the minimal condition of the potential, the $3\times3$ Majorana neutrino mass matrix is generated as
\begin{eqnarray}
m_{\nu_{ij}} = Y_{ij}\frac{\mu v^2}{M^2_{\Delta}} = Y_{ij}v_{\Delta}.
\end{eqnarray}
Here $v$ and $v_{\Delta}$ are the vacuum expectation values of standard model Higgs and the triplet $\Delta$ respectively. We continue the idea that the type-II seesaw scalar triplet $\Delta$ as the inflaton for chaotic inflation~\cite{Chen:2010uc} and combine the proposal in Ref.~\cite{Wetterich:2007kr} that the mass of $\Delta$ depends on the field value of cosmon field $\sigma$ is assumed in the following way,
\begin{eqnarray}\label{varytriplet}
M^2_{\Delta} = c_{\Delta}M^2_{GUT}\left[ 1 - \frac{1}{\tau}{\rm exp}(-\epsilon\frac{\sigma}{M_P})\right].
\label{eq11}
\end{eqnarray}
Here $M_{GUT}$ is the grand unification scale, and $c_{\Delta}$ and $\tau$ are the order one parameters. The potential of the cosmon field is given by
\begin{equation}
V(\sigma)=M_P^4 e^{-\alpha \sigma /M_P}
\label{eqa12}
\end{equation}
 with $\alpha \gae 10$ (from early dark energy constraint \cite{Doran:2007ep}). This term will result in a tracker behavior of the cosmon field after inflation. 
The evolution of $\sigma$ is given by \cite{Wetterich:1987fk}
\begin{equation}
\ddot{\sigma} + 3H\dot{\sigma} = -\frac{\partial V}{\partial\sigma} + \frac{\beta(\sigma)}{M}(\rho_{\nu}-3p_\nu)
\end{equation}
where
\begin{equation}
\beta(\sigma) \equiv \frac{\frac{\epsilon}{\tau}{\rm exp}(-\epsilon\frac{\sigma}{M_P})}{1 - \frac{1}{\tau}{\rm exp}(-\epsilon\frac{\sigma}{M_P})}
\end{equation}
We can write $\beta$ in the form $\beta(\sigma) = \frac{M_P}{\sigma - \sigma_t}$ for $\sigma_t \equiv-M_P\frac{\ln{\tau}}{\epsilon} $ when $\sigma$ is close to $\sigma_t$.
When $\sigma$ approaches $\sigma_t$ Eq.~(\ref{eqa12}) would behave like a cosmological constant and becomes the dark energy. For $\alpha \sigma_t / M_P \sim 276$ the cosmological constant has a value compatible with observation. This implies $\epsilon=-\alpha \ln \tau/276$ therefore if we choose $\alpha=10$ and $\ln \tau=1$, we would have $\epsilon=-0.05$ which implies a mild dependence of $M_\Delta$ on $\sigma$ through Eq.~(\ref{varytriplet}). Furthermore it is pointed out in Ref.~\cite{Wetterich:2007kr} that the detail form of $\sigma$-dependence of $M_{\Delta}$ is not important as long as a Taylor expansion is applicable around $\sigma \approx \sigma_{t}$\footnote{A possible derivation of the exponential mass dependence on the cosmon field associated with supersymmetry breaking is obtained in \cite{Chen:2011dv}.}. We will use those values in the following context.

Since $\Delta$ is the inflaton and can decay into several channels such as $\Delta \rightarrow \nu\nu, HH, ZZ,$ and $\sigma\sigma$ to reheat the universe. One should note that the decay mode of $\Delta \rightarrow \sigma\sigma$ is possible via the nonrenormalizable couplings which are from the expansion of Eq.~(\ref{varytriplet}), read
\begin{eqnarray}\label{cosmondecay}
L_{eff} = -\frac{c_{\Delta\epsilon^2}}{4\tau}(\frac{M_{GUT}}{M_{P}})^2\sigma^2\Delta^2.
\end{eqnarray}
The main decay widths of $\Delta$ are given by
\begin{eqnarray}\label{decaywidths1}
\Gamma_{\Delta}(\nu_{i}\nu_{j}) &=& \frac{Y^2_{ij}}{8\pi(1 + \delta_{ij})}M_{\Delta}, \\
\Gamma_{\Delta}(HH) &=& \frac{M^3_{\Delta}v^2_{\Delta}}{8\pi v^4}, \\
\Gamma_{\Delta}(ZZ) &=& \frac{g^2m_{Z}v^2_{\Delta}}{4\pi M_{\Delta}\cos^2{\theta_{W}}v^2},
\end{eqnarray}
and
\begin{eqnarray}\label{decaywidth2}
\Gamma_{\Delta}(\sigma\sigma) = \frac{c^2_{\Delta}\epsilon^4}{256\pi\tau^2}(\frac{M_{GUT}}{M_{P}})^4\frac{\rho_{\Delta}}{M^3_{\Delta}}.
\end{eqnarray}
The decay product produced through the coupling in Eq.~(\ref{cosmondecay}) might explain the hint of the need for an extra, dark, relativistic energy component in recent analyses~\cite{Komatsu:2010fb,Dunkley:2010ge,Hamann:2010bk,Mangano:2006ur,Seljak:2006bg,Izotov:2010ca}. The energy density of the new degree of freedom is usually normalized to neutrino energy density $\rho_{\nu}$ in a convenient way with the ``effective number of equivalent neutrinos" $N_{\nu eff}$ defined by
\begin{eqnarray}
\rho_{\nu} = \rho_{\gamma}\frac{7}{8}(\frac{4}{11})^{4/3}N_{\nu eff}.
\end{eqnarray}
$N_{\nu eff} = 4.6\pm0.8$ at $68\%$ C.L. for the experimental results of WMAP + BAO (baryon acoustic oscillations) + $H_{0}$ (the Hubble constant)~\cite{Dunkley:2010ge}, and the current observed primordial Helium mass fraction prefers a larger value $Y_{p} = 0.2565\pm0.0010(\rm stat.)\pm0.0050(\rm syst.)$ than standard BBN prediction $Y_{p} = 0.2487\pm0.0002$~\cite{Izotov:2010ca}. 
We take the constraint that the energy density of cosmon field is smaller than that of neutrinos during Big Bang Nucleosynthesis (BBN) epoch
\begin{eqnarray}\label{BBNconstraint}
\frac{\rho_{\sigma}}{s}|_{BBN} \lae \frac{\rho_{\nu}}{s}|_{BBN}.
\end{eqnarray}
Here $s$ is the entropy density and $\rho_{\sigma}|_{decay} = Br(\Delta\rightarrow2\sigma)M_{\Delta}n_{\Delta}$ with $Br$ and $n_{\Delta}$ denote branching ratio of and number density of inflaton $\Delta$. And since $\frac{n_{\Delta}}{s} = \frac{3}{4}\frac{T_{R}}{M_{\Delta}}$ we have
\begin{eqnarray}
\frac{\rho_{\sigma}}{s}|_{BBN} = \frac{\rho_{\sigma}}{s}|_{T_{R}}(\frac{T_{BBN}}{T_{R}}) = \frac{3}{4}Br(\Delta\rightarrow2\sigma)T_{BBN}
\end{eqnarray}
and
\begin{eqnarray}
\frac{\rho_{\nu}}{s}|_{BBN} \sim \frac{3\times\frac{3}{4}T_{BBN}}{19}
\end{eqnarray}
for the two sides of Eq.~(\ref{BBNconstraint}). Therefore we roughly have the constraint for the $Br(\Delta\rightarrow\sigma\sigma) \lae {\cal O}(\frac{1}{10})$. The bound can be easily satisfied in Eqs.~(\ref{decaywidths1}) - (\ref{decaywidth2}). It also can be understood that the decay rate is proportional to the Hubble parameter squared and decreases faster than the universe expansion rate. Inflaton will not decay completely into radiation and reheat the universe if the four-point interaction (Eq.~(\ref{cosmondecay})) is the dominant decay.


\underline{\emph{Primordial density perturbations}}: 
If the cosmon $\sigma$ is light during inflation, it is subject to fluctuations similar to the inflaton, namely, $\delta \sigma \sim H_\ast/2\pi$. This would lead to fluctuations of the decay width by the variation of the inflaton mass $M_{\Delta}$, and may contribute to the primordial density perturbation. The potential of the cosmon field is given by Eq.~(\ref{eq5}) and (\ref{eq11}):
\begin{equation}
V(\sigma)=\frac{1}{2} c_{\Delta}M^2_{GUT}\left[ 1 - \frac{1}{\tau}{\rm exp}(-\epsilon\frac{\sigma}{M_P})\right]\Delta^2.
\label{eq22}
\end{equation}
There is another term in the potential given by Eq.~(\ref{eqa12}) but it is subdominant and negligible (for a wide range of $\sigma$) during inflation. 

The condition of the cosmon being ``light'' (during inflation) is given by
\begin{equation}
|V^{\prime\prime}/H^2| \sim |\epsilon \beta| \ll 1
\label{eq23}
\end{equation}
 Because $|\epsilon| \sim 0.01$, Eq.~(\ref{eq23}) can be satisfied if $|\beta| \lae 100$.

The primordial curvature perturbation can be obtained by using Eq.~(\ref{eq7}) and (\ref{eq11}):
\begin{equation}
\zeta \sim \frac{\delta \Gamma}{\Gamma} \sim \frac{\delta M_{\Delta}}{M_{\Delta}}=|\beta_{end}|\frac{\delta\sigma}{M_P} \sim 10^{-5}.
\label{eq14}
\end{equation}
Here $|\beta_{end}|$ means $|\beta|$ at the end of inflation.
We could see that for $|\beta_{end}| \gae 1$ we have $\frac{\delta\sigma}{M_P} \sim H_\ast/M_P \lae 10^{-5}$. Note that the condition that $\zeta_{inf}$ subdominant would require $|\beta_{end}| \gae 1$. 
Therefore we require $1 \lae |\beta_{end}| \lae 10^2$. In this case, the primordial density perturbation is dominated by the fluctuation of the cosmon field which would play the role of dark energy in the current universe. 

Actually we can also consider $|\beta_{end}| \lae 1$. In this case, $\sigma$ would be slow-rolling until inflation ends. According to Eq.~(\ref{eq14}), the contribution of primordial curvature perturbation is subdominant. However, it is still possible to generate sizeable non-Gaussianity \cite{Zaldarriaga:2003my,Ichikawa:2008ne}. We will discuss this in the following section.


\underline{\emph{Non-Gaussianity}}:
From Eq.~(\ref{eq10}) we can obtain
\begin{eqnarray}
\frac{6}{5}f_{NL} = 6(1 + {\cal O}(1)\frac{\epsilon}{\beta}),
\end{eqnarray}
where the order one factor depends on different decay widths in Eqs.~(\ref{decaywidths1}) - (\ref{decaywidth2}). From here we can see that larger $\beta$ implies smaller $f_{NL}$. This may be intuitively understood by the following argument. If we require $\zeta \propto \beta \delta\sigma \sim 10^{-5}$, large $\beta$ implies small $\delta \sigma$ which means the nonlinear (non-Gaussian) effect $\propto (\delta \sigma)^2$ is small. In the case where the contribution of the primordial density perturbation is dominated by the fluctuation of the cosmon field, we have $|\epsilon| \sim 0.01$ and $|\beta| \gae 1$ therefore $f_{NL}=5$ which may be detected in the near future by PLANCK satellite.

If the contribution of $\zeta$ from modulated reheating is subdominant, from
Eq.~(\ref{eq3}), it can be shown that $f_{NL}$ would be reduced by a factor of $\beta^2/(1+\beta)^2$. For example, if $\beta \sim 0.5 $, we would have $f_{NL} \sim {\cal O}(1)$ which is close to the marginal value of experimental sensitivity in the near future. The non-Gaussianity produced is still larger than the case that we only have chaotic inflation without modulated reheating.

\underline{\emph{Isocurvature and leptogenesis}}: As we have shown in~\cite{Chen:2010uc} the baryon asymmetry of the universe can be obtained via leptogenesis if two triplet scalars exist. 
In our model, if the primordial density perturbation is dominated by the flucturation of the cosmon field through modulated reheating, it is possible to generate a large baryonic isocurvature perturbation. 
Let's consider the possibility of isocurvature perturbation induced from the lepton asymmetry in this construction. 

The $CP$ violation is generated through the interference between the tree level and self-energy correction of the triplet scalar decay, given by 
\begin{eqnarray}
\epsilon_{1} &\approx& \frac{\rm{Im}[\mu_{1}\mu^*_{2}\sum_{k,l}(Y_{1kl}Y^*_{2kl})]}{8\pi^2(M^2_{\Delta_1} - M^2_{\Delta_2})}\Big(\frac{M_{\Delta_{1}}}{\Gamma_{\Delta_1}}\Big).
\end{eqnarray}
$\mu_{1,2}$ are the cubic couplings involving the triplet and two powers of the Higgs-doublet and indices $1,2$ represent the physical quantities refer to the two triplets scalars $\Delta_{1,2}$. The parameter $K$ is defined by $K = \Gamma_{\Delta_{1}}/H(T = M_{\Delta_{1}})$ with $ H(T)|_{T=M_{\Delta_{1}}} = \sqrt{\frac{4\pi^3g_{*}}{45}}\frac{M_{\Delta_{1}}^2}{M_{P}}$ (here we assume $M_{\Delta_{1}} < M_{\Delta_{2}}$) and $g_{*}\sim100$ is the effective number of massless particles. After solving the Boltzmann equations that involve decay, inverse decay, and annihilation processes, the baryon asymmetry can be approximated by 
\begin{eqnarray}
\frac{n_{B}}{s} \sim 0.5\times10^{-2}\epsilon_{1}\times(K^2 + 9)^{-1/2}
\end{eqnarray}

for $0 < K < 10$~\cite{earlyuniverse}. Let $M_{\Delta_{2}} = 3\times10^{13}$ GeV, $\mu_{1,2} \sim 10^{12}$ GeV, $Y_{(1,2)ij}\sim0.1$ with $m_{\nu} \sim 10^{-1\sim-2}$ eV, and $K = 5$, the $n_{B}/s \approx 10^{-10}$ as desired. We take $\epsilon_{1} \propto M^{-2}_{\Delta_{1}}$ and $K \propto M^{P}_{\Delta_{1}}$ where $P$ is integer and depends on the decay widths given in Eqs.~(\ref{decaywidths1}) - (\ref{decaywidth2}). The baryon-isocurvature fluctuation can be expressed as     
\begin{equation}
S_{B} \equiv \frac{\delta (n_b/s)}{n_b/s}=\zeta\left[-1 - \frac{P}{2}K^2(K^2 + 9)^{-1}\right].
\end{equation}
\begin{figure}[t]
  \centering
\includegraphics[width=0.4\textwidth]{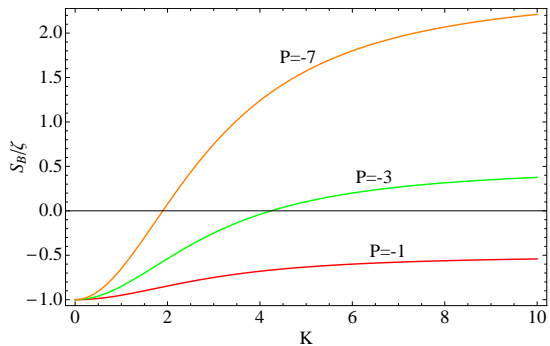}
  \caption{The ratio of isocurvature to primordial density perturbations versus the parameter $K$ and $P = -1, -3, -7$ correspond to $\nu\nu, HH, ZZ$ decay modes respectively.}
  \label{fig:isocurvature}
\end{figure} 
The observational constraint on the uncorrelated baryon isocurvature is $|S_{B}/\zeta| \lae {\cal O}(1)$~\cite{Komatsu:2009kd,Komatsu:2010fb}. We show the contributions to $S_{B}$ from the decay modes\footnote{We ignore the $\sigma\sigma$ mode as it is supposed to be subdominant.} of $\Delta_{1}$ as the function of $K$ in Fig.~\ref{fig:isocurvature}. It indicates the model is well within constraint. For the case of primordial density perturbations from modulated reheating is subdominant we expect the $S_{B}$ is smaller.  

\underline{\emph{Conclusion}}:  In this paper, we investigated the possible cosmological consequences of inflation driven by a Higgs triplet of type II seesaw mechanism if the dark energy is from the growing neutrino mechanism. Interestingly, in this setup, we found that the primordial curvature perturbation and/or non-Guassianity may be from the quantum fluctuations of the cosmon field which would cause the dark energy we observe today and cosmon would play the role of the modulon. If the contribution of the modulated reheating to the curvature perturbation dominates, there is an issue of baryon isocurvature perturbation. In this case, we investigated the allowed region of the parameter space and found the constraint is not very severe. Futhermore, if isocurvature perturbation is found in future experiments, it would provide an interesting constraint to our model. If the curvature perturbation from the modulon is subdominate, there is no issue about isocurvature perturbation. However, sizable non-Gaussianity may still be generated. \\

\acknowledgments
CML was supported by the NSC under grant No. NSC 99-2811-M-007-068 and 
CSC was supported by the National Center of Theoretical Sciences of Taiwan (NCTS).
We thank Kazunori Kohri for useful discussion. CSC would like to thank KEK for hospitality.


\begin{thebibliography}{99}

\bibitem{Lyth:2009zz}
  For a review, see e.g.,~D.~H.~Lyth, A.~R.~Liddle,
  Cambridge, UK: Cambridge Univ. Pr. (2009) 497 p.

\bibitem{Sasaki:1995aw}
  M.~Sasaki, E.~D.~Stewart,
  Prog.\ Theor.\ Phys.\  {\bf 95}, 71-78 (1996).
  [astro-ph/9507001].

\bibitem{Sasaki:1998ug}
  M.~Sasaki, T.~Tanaka,
  Prog.\ Theor.\ Phys.\  {\bf 99}, 763-782 (1998).
  [gr-qc/9801017].

\bibitem{Lyth:2004gb}
  D.~H.~Lyth, K.~A.~Malik, M.~Sasaki,
  JCAP {\bf 0505}, 004 (2005).
  [astro-ph/0411220].

\bibitem{Lyth:2005fi}
  D.~H.~Lyth, Y.~Rodriguez,
  Phys.\ Rev.\ Lett.\  {\bf 95}, 121302 (2005).
  [astro-ph/0504045].

\bibitem{Maldacena:2002vr}
  J.~M.~Maldacena,
  JHEP {\bf 0305}, 013 (2003).
  [astro-ph/0210603].


\bibitem{Chen:2010uc}
  C.~-S.~Chen, C.~-M.~Lin,
  Phys.\ Lett.\  {\bf B695}, 9-12 (2011).
  [arXiv:1009.5727 [hep-ph]].

\bibitem{Linde:2007fr}
  A.~D.~Linde,
  Lect.\ Notes Phys.\  {\bf 738}, 1-54 (2008).
  [arXiv:0705.0164 [hep-th]].


\bibitem{Dvali:2003em}
  G.~Dvali, A.~Gruzinov, M.~Zaldarriaga,
  Phys.\ Rev.\  {\bf D69}, 023505 (2004).
  [astro-ph/0303591]; L.~Kofman,
  arXiv:astro-ph/0303614.


\bibitem{Amendola:2007yx}
  L.~Amendola, M.~Baldi and C.~Wetterich,
  Phys.\ Rev.\  D {\bf 78}, 023015 (2008)
  [arXiv:0706.3064 [astro-ph]].

\bibitem{Wetterich:2007kr}
  C.~Wetterich,
  Phys.\ Lett.\  B {\bf 655}, 201 (2007)
  [arXiv:0706.4427 [hep-ph]].

\bibitem{Chen:2011dv}
  C.~-S.~Chen, C.~-M.~Lin,
  Eur.\ Phys.\ J.\  {\bf C71}, 1643 (2011).
  [arXiv:1101.4362 [hep-ph]].
  
\bibitem{type-II} W.~Konetschny and W.~Kummer,
  Phys.\ Lett.\  B {\bf 70}, 433 (1977); J.~Schechter and J.~W.~F.~Valle,
  Phys.\ Rev.\  D {\bf 22}, 2227 (1980); T.~P.~Cheng and L.~F.~Li,
  Phys.\ Rev.\  D {\bf 22}, 2860 (1980); M.~Magg and C.~Wetterich,
  Phys.\ Lett.\  B {\bf 94}, 61 (1980); G.~Lazarides, Q.~Shafi and C.~Wetterich,
  Nucl.\ Phys.\  B {\bf 181}, 287 (1981); G.~B.~Gelmini and M.~Roncadelli,
  Phys.\ Lett.\  B {\bf 99}, 411 (1981); R.~N.~Mohapatra and G.~Senjanovic,
  Phys.\ Rev.\  D {\bf 23}, 165 (1981).



\bibitem{Doran:2007ep}
  M.~Doran, G.~Robbers, C.~Wetterich,
  Phys.\ Rev.\  {\bf D75}, 023003 (2007).
  [astro-ph/0609814].

\bibitem{Wetterich:1987fk}
  C.~Wetterich,
  Nucl.\ Phys.\  {\bf B302}, 645 (1988); P.~Gu, X.~Wang, X.~Zhang,
  Phys.\ Rev.\  {\bf D68}, 087301 (2003).
  [hep-ph/0307148]; R.~Fardon, A.~E.~Nelson, N.~Weiner,
  JCAP {\bf 0410}, 005 (2004).
  [astro-ph/0309800]; A.~W.~Brookfield, C.~van de Bruck, D.~F.~Mota, D.~Tocchini-Valentini,
  Phys.\ Rev.\ Lett.\  {\bf 96}, 061301 (2006).
  [astro-ph/0503349]; N.~Afshordi, M.~Zaldarriaga, K.~Kohri,
  Phys.\ Rev.\  {\bf D72}, 065024 (2005).
  [astro-ph/0506663]; O.~E.~Bjaelde, A.~W.~Brookfield, C.~van de Bruck, S.~Hannestad, D.~F.~Mota, L.~Schrempp, D.~Tocchini-Valentini,
  JCAP {\bf 0801}, 026 (2008).
  [arXiv:0705.2018 [astro-ph]]; K.~Ichiki, Y.~-Y.~Keum,
  JCAP {\bf 0806}, 005 (2008).
  [arXiv:0705.2134 [astro-ph]]; R.~Takahashi, M.~Tanimoto,
  Phys.\ Lett.\  {\bf B633}, 675-680 (2006).
  [hep-ph/0507142]; R.~Takahashi, M.~Tanimoto,
  JHEP {\bf 0605}, 021 (2006).
  [astro-ph/0601119].





\bibitem{Dunkley:2010ge}
  J.~Dunkley {\it et al.},
  arXiv:1009.0866 [astro-ph.CO].

\bibitem{Hamann:2010bk}
  J.~Hamann, S.~Hannestad, G.~G.~Raffelt, I.~Tamborra and Y.~Y.~Y.~Wong,
  Phys.\ Rev.\ Lett.\  {\bf 105}, 181301 (2010)
  [arXiv:1006.5276 [hep-ph]].

\bibitem{Mangano:2006ur}
  G.~Mangano, A.~Melchiorri, O.~Mena, G.~Miele and A.~Slosar,
  JCAP {\bf 0703}, 006 (2007)
  [arXiv:astro-ph/0612150].

\bibitem{Seljak:2006bg}
  U.~Seljak, A.~Slosar and P.~McDonald,
  JCAP {\bf 0610}, 014 (2006)
  [arXiv:astro-ph/0604335].





\bibitem{Izotov:2010ca}
  Y.~I.~Izotov and T.~X.~Thuan,
  Astrophys.\ J.\  {\bf 710}, L67 (2010)
  [arXiv:1001.4440 [astro-ph.CO]].

 \bibitem{Komatsu:2010fb}
  E.~Komatsu {\it et al.}  [WMAP Collaboration],
  Astrophys.\ J.\ Suppl.\  {\bf 192}, 18 (2011)
  [arXiv:1001.4538 [astro-ph.CO]].



\bibitem{Zaldarriaga:2003my}
  M.~Zaldarriaga,
  Phys.\ Rev.\  {\bf D69}, 043508 (2004).
  
\bibitem{Ichikawa:2008ne}
  K.~Ichikawa, T.~Suyama, T.~Takahashi, M.~Yamaguchi,
  Phys.\ Rev.\  {\bf D78}, 063545 (2008).
  [arXiv:0807.3988 [astro-ph]].
  
\bibitem{earlyuniverse} E.W.~Kolb, M.S.~Turner, The Early Universe, Addison-Wesley, Reading, MA, 1990.  

\bibitem{Komatsu:2009kd}
  E.~Komatsu {\it et al.},
  arXiv:0902.4759 [astro-ph.CO].
  

 


\end{thebibliography}
\end{document}